\documentclass{webofc}
\usepackage[varg]{txfonts}   

\usepackage{psfrag} 

\wocname{EPJ Web of Conferences}
\woctitle{QCD@Work 2014}

\newcommand{\ket}{\,\rangle}
\newcommand{\bra}{\langle \,}
\def\Comment#1{}
\newcommand{\bean}{\begin{eqnarray*}}
\newcommand{\eean}{\end{eqnarray*}}

\newcommand{\gapproxeq}{\lower
.7ex\hbox{$\;\stackrel{\textstyle >}{\sim}\;$}}
\newcommand{\lapproxeq}{\lower
.7ex\hbox{$\;\stackrel{\textstyle <}{\sim}\;$}}

\newcommand\lsim{\mathrel{\rlap{\lower4pt\hbox{\hskip1pt$\sim$}}
    \raise1pt\hbox{$<$}}}
\newcommand\gsim{\mathrel{\rlap{\lower4pt\hbox{\hskip1pt$\sim$}}
    \raise1pt\hbox{$>$}}}
\newcommand{\ba}{\begin{array}}
\newcommand{\ea}{\end{array}}
\newcommand{\nn}{\nonumber}

\newcommand{\be}{\begin{equation}}
\newcommand{\eequ}{\end{equation}}
\newcommand{\bear}{\begin{eqnarray}}
\newcommand{\eear}{\end{eqnarray}}
\newcommand{\tab}{\hspace*{0.5cm}}

\newcommand{\cO}{{\cal O}}

\newcommand{\mL}{\mathcal{L}}

\newcommand{\mM}{\mathcal{M}}

\newcommand{\Frac}[2]{\frac{\displaystyle #1}{\displaystyle #2}}
\newcommand{\Int}{\displaystyle{\int}}

\begin{document}
%
\hspace*{10.5cm}{\small   {IFT-UAM/CSIC-14-096}  }
\\
\hspace*{11cm}{\small {FTUAM-14-37} }
\\

\title{\vspace*{-1.5cm} Electroweak chiral Lagrangians \\
and the Higgs properties at the one-loop level}
%
%

\author{
\vspace*{-0.25cm}
J.J. Sanz-Cillero\inst{1}\fnsep\thanks{
\email{juanj.sanz@uam.es}
\\
I would like to thank the organizers for their work and the lively discussion during the
workshop; also for their patience.
This work is partially supported by
the Spanish Government  and ERDF funds from the European Commission
[FPA2010-17747,
FPA2013-44773-P,   
SEV-2012-0249,
CSD2007-00042] and the Comunidad de Madrid [HEPHACOS    S2009/ESP-1473].
}
}

\institute{
Departamento de F\'isica Te\'orica and Instituto de F\'isica Te\'orica,
IFT-UAM/CSIC\\
Universidad Aut\'onoma de Madrid,  C/ Nicol\'as Cabrera 13-15, \\
Cantoblanco, 28049 Madrid, Spain
          }

\abstract{%
In these proceedings we explore the use of (non-linear) electroweak chiral Lagrangians
for the description of possible beyond the Standard Model (BSM) strong dynamics in the electroweak (EW) sector.
Experimentally one observes an approximate  EW symmetry breaking pattern
$SU(2)_L\times SU(2)_R/SU(2)_{L+R}$.
Quantum Chromodynamics (QCD) shows a similar chiral structure~\cite{chpt}
and, in spite of the
differences (in the EW theory $SU(2)_L\times U(1)_Y$ is gauged),
it has served for years as a guide for this type of studies~\cite{Morales:94,Appelquist:1980vg,Longhitano:1980iz}.
Examples of one-loop computations in the low-energy effective theory and the theory including the first vector (V) and axial-vector (A) resonances
are provided, yielding, respectively, predictions for $\gamma\gamma\to Z_LZ_L,W^+_LW^-_L$ and the oblique parameters $S$ and $T$.
}
\maketitle

\vspace*{-0cm}
\section{Introduction: strong dynamics and chiral Lagrangians}

\tab A non-linear realization of the EW would-be Goldstone bosons (WBGBs) is considered
to build the EW low-energy effective field theory (EFT),
which is described by an EW chiral Lagrangian with a light Higgs (ECLh).
It includes the Standard Model (SM) content:
the EW Goldstones $w^a$, the EW gauge bosons $W^a_\mu$ and $B_\mu$ and a singlet Higgs $h$
(the fermion sector is not discussed here).
In particular, in Sec.~\ref{sec.ECLh} we explain the chiral counting
in the ECLh~\cite{photon-scat,EW-chiral-counting}
and provide an example of a next-to-leading order (NLO)  computation:
we calculate $\gamma\gamma\to W^+_L W^-_L,\, Z_L Z_L$  within this framework
up to the one-loop level~\cite{photon-scat} at energies below new possible composite resonances,
$\sqrt{s}\ll \Lambda_{\rm ECLh} \sim $min$\{M_R ,\, 4\pi v\}$
(with $v=(\sqrt{2} G_F)^{-1/2}$ and $4\pi v\simeq 3$~TeV).
Analogous works on $WW$--scattering can be found in Refs.~\cite{WW-theory}.

However, in the case of having heavy composite resonance,
the EFT stops being valid when the energy becomes of the order of their masses
(expected to be of the order of $M_R\sim 4\pi v\sim 3$~TeV).
One has to introduce these new degrees of freedom
in our EW Lagrangian
following a procedure analogous to that in QCD~\cite{RChT}.
Likewise, under reasonable ultraviolet (UV) completion hypotheses like,
e.g., the Weinberg sum-rules (WSRs) fulfilled by certain types
of theories~\cite{Bernard:1975cd,Peskin:92,WSR,S-Orgogozo:11,WalkingTC},  one can make predictions
on low-energy observables. In Sec.~\ref{sec.oblique}
we write down the relevant $SU(2)_L\times SU(2)_R$ invariant Lagrangian
including the SM content and a multiplet of V and A resonances and extract
one-loop limits on the resonance masses and the Higgs coupling $g_{hWW}$~\cite{S+T+scalar}
from the experimental values of oblique parameters $S$ and $T$~\cite{S+T-phenomenology}.
Alternative one-loop analyses
can be found in Refs.~\cite{oblique-other,S-Orgogozo:11}.

\section{Low-energy EFT: ECLh and one-loop $\gamma\gamma\to W_L^a W_L^b$ scattering}
\label{sec.ECLh}

The Higgs boson does not enter in the SM at tree-level in these processes (where one also
has ${  \mM(\gamma\gamma\to ZZ)^{\rm tree}_{\rm SM}=0  }$). Nevertheless, one can search
for new physics by studying the one-loop corrections~\cite{photon-scat},
which are sensitive to deviations from the SM in the Higgs boson couplings.
Our analysis~\cite{photon-scat} is performed in the Landau gauge and making use of the
Equivalence Theorem (Eq.Th.)~\cite{equivalence-theorem},
\bear
\mM(\gamma\gamma\to W_L^a W_L^b) &\simeq& \, -\, \mM(\gamma\gamma\to w^a w^b) \, ,
\eear
valid in the energy regime $m_{W}^2, m_{Z}^2\ll s$.
The EW gauge boson masses $m_{W,Z}$ are then neglected
in our computation.
Furthermore, since $m_h\sim m_{W,Z}\ll 4\pi v\simeq 3$~TeV
we also neglect $m_h$ in our calculation.
In summary, the applicability range in~\cite{photon-scat} is
\bear
m_W^2,\, m_Z^2 ,\, m_h^2 \quad \stackrel{\rm Eq.Th.}{\ll} \quad
s,\, t,\, u \quad \stackrel{\rm EFT}{\ll}  \quad \Lambda_{\rm ECLh}^2 \, ,
\eear
with the upper limit given by the  EFT cut-off  $\Lambda_{\rm ECLh}$,
expected to be of the order  of $4\pi v\simeq 3$~TeV or the mass of possible heavy BSM particles.

The WBGBs are described by a matrix field $U$ that takes values
in the $SU(2)_L \times SU(2)_R/SU(2)_{L+R}$ coset,   and transforms as
$U \to L U R^\dagger$~\cite{Appelquist:1980vg,Longhitano:1980iz}.
The relevant ECLh  with the basic building blocks is
\begin{eqnarray}
&&
\hspace*{-1.5cm}
U=u^2=1 +i w^a\tau^a/v+\cO(w^2)\, , \;
D_\mu U =\partial_\mu U + i\hat{W}_\mu U - i U\hat{B}_\mu \, ,
\;
V_\mu = (D_\mu U) U^\dagger \, ,\;
\;
u^\mu = -i\, u^\dagger  D^\mu U\, u^\dagger\, ,
\label{VmuandT}
\label{eq.cov-deriv}
\nn \\
&&
\hspace*{-1.5cm}
\hat{W}_{\mu\nu} = \partial_\mu \hat{W}_\nu - \partial_\nu \hat{W}_\mu
 + i  [\hat{W}_\mu,\hat{W}_\nu ],
\quad 
\hat{B}_{\mu\nu}  = \partial_\mu \hat{B}_\nu -\partial_\nu \hat{B}_\mu ,
\label{fieldstrength}
\quad \hat{W}_\mu = g W_\mu^a  \tau^a/2 ,\;\hat{B}_\mu = g'\, B_\mu \tau^3/2 \, ,
\label{EWfields}
\end{eqnarray}
with well-defined trasnformation properties~\cite{photon-scat,Longhitano:1980iz,S+T+scalar}.
Two particular parametrizations of the unitary matrix $U$ (exponential and spherical)
were considered in~\cite{photon-scat}, both leading to the
same predictions for the physical (on-shell) observables.~\footnote{
Other representations have been recently studied in Ref.~\cite{Machado:2014}.}
We consider the counting
 $\partial_\mu\, ,m_W\, , m_Z, m_h \sim \cO(p) $,
$D_\mu U,\; V_\mu\sim \cO(p)$ and
$\hat{W}_{\mu\nu},\;\hat{B}_{\mu\nu} \sim  \cO(p^2)$~\cite{photon-scat,EW-chiral-counting}.
%
We require the ECLh Lagrangian to be CP invariant,
Lorentz invariant and $SU(2)_L \times U(1)_Y$ gauge invariant.
Here we focus ourselves on the relevant terms for
$\gamma \gamma \to w^a w^b$
at leading order (LO) --$\cO(p^2)$--
and NLO in the chiral counting --$\cO(p^4)$--~\cite{photon-scat,Longhitano:1980iz}:
\bear
&& \hspace*{-1.25cm}
\mL_2 =    -\Frac{1}{2 g^2} \bra \hat{W}_{\mu\nu}\hat{W}^{\mu\nu}\ket
-\Frac{1}{2 g^{'2}} \bra \hat{B}_{\mu\nu} \hat{B}^{\mu\nu}\ket
+\Frac{v^2}{4}\left[%
  1 + 2a \Frac{h}{v} + b \Frac{h^2}{v^2}\right] \bra D^\mu U^\dagger D_\mu U \ket
 + \Frac{1}{2} \partial^\mu h \, \partial_\mu h
 + \dots\, ,
\\
&& \hspace*{-1.25cm}
\mL_4 =
  a_1 {\rm Tr}(U \hat{B}_{\mu\nu} U^\dagger \hat{W}^{\mu\nu})
  + i a_2 {\rm Tr} (U \hat{B}_{\mu\nu} U^\dagger [V^\mu, V^\nu ])
  - i a_3  {\rm Tr} (\hat{W}_{\mu\nu}[V^\mu, V^\nu])
 -\Frac{c_{\gamma}}{2}\Frac{h}{v}e^2 A_{\mu\nu} A^{\mu\nu}\, +\, ...
\label{eq.L4}
\nn
\eear
where  $\langle X \rangle$ stands for the trace of the $2\times 2$ matrix $X$, one has
the photon field strength  $A_{\mu \nu}=\partial_\mu A_\nu- \partial_\nu A_\mu$
and the dots stand for operators not relevant
within our approximations for  $\gamma\gamma$-scattering~\cite{photon-scat}.
%
%

The amplitudes $\mM(\gamma(k_1,\epsilon_1)\gamma(k_2,\epsilon_2)\to w^a(p_1) w^b(p_2))$,
with $w^a w^b=zz,w^+w^-$, have the structure
\be
\mM   =
ie^2 (\epsilon_1^\mu \epsilon_2^\nu T_{\mu\nu}^{(1)}) A(s,t,u)+
ie^2 (\epsilon_1^\mu \epsilon_2^\nu T_{\mu\nu}^{(2)})B(s,t,u),
\eequ
written in terms of the two independent Lorentz structures
$T_{\mu\nu}^{(1,2)}\sim\cO(p^2)$
involving the external momenta, which can be found in~\cite{photon-scat}.
The Mandelstam variables are defined as
$s=(p_1+p_2)^2$, $t=(k_1-p_1)^2$ and $u=(k_1-p_2)^2$ and the $\epsilon_i$'s
are the  polarization vectors of the external photons.

In dimensional regularization, our NLO computation of  the $\mM(\gamma\gamma\to w^a w^b)$
amplitudes can be systematically  sorted out in the form~\cite{photon-scat}
\vspace*{-0.5cm}
\begin{equation}
\mM=\mM_{\rm LO}+\mM_{\rm NLO}
\quad \sim\quad
\underbrace{\cO(e^2)}_{\rm LO,\, tree}
\quad +\quad \left(
\underbrace{\cO\left( e^2\Frac{p^2}{16\pi^2 v^2}\right) }_{\rm NLO,\, 1-loop}
\quad+\quad
\underbrace{\cO\left( e^2 \Frac{a_i p^2}{v^2}\right) }_{\rm NLO,\, tree}
 \right) \,  ,
 \label{eq.M}
\end{equation}
where $e\sim \cO(p/v)$
and  $A$ and $B$ are given up to NLO by~\cite{photon-scat}
\bear
A(\gamma\gamma\to zz)_{\rm LO}
&=& B(\gamma\gamma\to zz)_{\rm  LO}
= 0 \, ,
\label{eq.photon-scat}
  \\
A(\gamma \gamma \to zz)_{\rm NLO}&=&
\Frac{2 a c_\gamma^r}{v^2} + \Frac{ (a^2-1)}{4\pi^2v^2}\, ,
\label{eq.A-zz}
\qquad\qquad\qquad  \qquad\qquad
B(\gamma \gamma \to zz)_{\rm NLO}= 0,
\label{eq.B-zz}
\nn \\
A(\gamma\gamma\to w^+ w^-)_{\rm LO}    
&=& 2 s B(\gamma\gamma\to w^+ w^-)_{\rm  LO}    
= -\Frac{1}{t} - \Frac{1}{u},
\nn\\
A(\gamma \gamma \to w^+w^-)_{\rm NLO}&=&
\Frac{2 a c_\gamma^r}{v^2} + \Frac{ (a^2-1)}{8\pi^2v^2}
+ \Frac{8(a_1^r-a_2^r+a_3^r)}{v^2}\, ,
\label{eq.A-zz}
\qquad
B(\gamma \gamma \to w^+w^-)_{\rm NLO}= 0.
\nn \label{eq.results}
\eear
The term with $c_\gamma^r$  comes from the Higgs tree-level
exchange in the $s$--channel, the term proportional to $(a^2-1)$
comes from the one-loop diagrams with $\mL_2$ vertices,
and the Higgsless operators in Eq.~(\ref{eq.L4}) yield the tree-level contribution
to $\gamma\gamma\to w^+ w^-$ proportional to $(a_1-a_2+a_3)$.
Independent diagrams are  in general UV divergent.
%
However, in dimensional regularization,
the final one-loop amplitude turns out to be UV finite
%
and one has $ a_1^r-a_2^r+a_3^r=a_1 -a_2 +a_3 $, $c^r_\gamma=c_\gamma$~\cite{photon-scat},
as in the Higgsless case~\cite{photon-scat-Higgsless}.

%

In order to pin down each of the relevant combinations of ECLh couplings
in Eq.~(\ref{eq.photon-scat})   ($a$, $c_\gamma^r$ and $a_1^r-a_2^r+a_3^r$)
one must combine our $\gamma\gamma$-scattering analysis with other observables that depend
on this same set of  parameters. It is not difficult to
find that other processes involving photons depend on these parameters.
In Ref.~\cite{photon-scat} we computed 4 more observables of this kind:
the $h\to \gamma\gamma$ decay width (depending on $a$ and $c_\gamma$),
the oblique $S$--parameter (depending on $a$ and $a_1$), and the
$\gamma^*\to w^+w^-$   (depending on $a$ and $a_2-a_3$)
and $\gamma^*\gamma \to h$ (depending on $c_\gamma$)
electromagnetic form-factors.
%
The one-loop contribution in these six relevant amplitudes is found to
be UV-divergent in some cases. These divergences are absorbed by means
of the generic $\cO(p^4)$ renormalizations $a_i^r(\mu)=a_i+\delta a_i$.
As expected, the renormalization in the six observables gives a fully consistent
set of renormalization conditions
and fixes the running of the renormalized couplings in the way
given in Table~\ref{tab.running}.

\begin{table}[!t]
\vspace*{-0.5cm}
\begin{center}
\begin{tabular}{ c|c|c }
 & {\bf ECLh  } &  {\bf ECL} (Higgsless)                 
\\[5pt] \hline
\rule{0pt}{3ex}
$\quad \Gamma_{a_1-a_2+a_3}\quad$ & $\quad 0 \quad $ &  0                
\\[5pt] \hline
\rule{0pt}{3ex}
$\quad \Gamma_{c_\gamma}\quad$ & $\quad 0 \quad $ &  -                
\\[5pt] \hline
\rule{0pt}{3ex}
$\quad \Gamma_{a_1}\quad$ & $\quad -\frac{1}{6}(1-a^2) \quad $&$\quad -\frac{1}{6} \quad $                
\\[5pt] \hline
\rule{0pt}{3ex}
$\quad \Gamma_{a_2-a_3} \quad$ & $\quad -\frac{1}{6}(1-a^2) \quad $& $\quad -\, \frac{1}{6} \quad $                  
\\[5pt] \hline
\rule{0pt}{3ex}
$\quad \Gamma_{a_4}\quad$ & $\quad \frac{1}{6}(1-a^2)^2  \quad $ & $\quad \frac{1}{6}\quad $                 
\\[5pt] \hline
\rule{0pt}{3ex}
$\quad \Gamma_{a_5}\quad$ & $\quad \frac{1}{8}(b-a^2)^2
+\frac{1}{12}(1-a^2)^2\quad $ &
$\quad \frac{1}{12}\quad $                 
\\[5pt] \hline
\end{tabular}
\caption{\small
Running
$\frac{da_i^r}{d\ln\mu} \,=\, - \Frac{\Gamma_{a_i}}{16\pi^2 }$
of the relevant ECLh parameters and their combinations appearing
in the six selected observables.
The third column provides the corresponding running
for the Higgsless EW chiral Lagrangian (ECL) case~\cite{Morales:94}.
For the sake of completeness,
we have added the running of the ECLh parameters $a_4^r$ and $a_5^r$,
which has been recently determined in the one-loop analysis of  $WW$--scattering
within the framework of chiral Lagrangians~\cite{WW-theory}.
One can see that in the SM limit ($a=b=1$) these $\mL_4$ coefficients  do not run,
in agreement with the fact that these higher order operators are absent in the SM.
}
\label{tab.running}
\end{center}
\end{table}


\section{Impact of spin--1 composite resonances on the oblique parameters}
\label{sec.oblique}

One can extend the range of validity and predictability
of the ECLh by adding possible new states to the theory.
Thus, the lightest V and A resonances are added
to the EW Lagrangian in Ref.~\cite{S+T+scalar}
in order to describe the oblique parameters $S$ and $T$~\cite{Peskin:92}.
The relevant EW chiral invariant Lagrangian is given by the kinetic and Yang-Mills terms and
the interactions~\cite{S+T+scalar}~\footnote{
%
Here we follow the notation
$f_\pm^{\mu\nu} =u^\dagger \hat{W}^{\mu\nu} u \pm u \hat{B}^{\mu\nu} u^\dagger$ from Ref.~\cite{S-Higgsless,S+T+scalar},
where there is a global sign difference with~\cite{photon-scat} in the definitions
of $\hat{W}_\mu$ and $\hat{B}_\mu$. The spin--1 resonances are described
in the antisymmetric tensor formalism~\cite{RChT}.
}$^{,}$~\footnote{
In other works, the coupling $a$ can be found with the notation $\kappa_W$ and  $\omega$~\cite{S+T+scalar}
or $\kappa_V$~\cite{LHC-exp}.}
\begin{eqnarray}\label{eq:Lagrangian}
\mathcal{L}\; &=\; &
\frac{v^2}{4}\,\bra \! u_\mu u^\mu \!\ket\,\left( 1 + \frac{2\,a }{v}\, h\right)
+ \frac{F_V}{2\sqrt{2}}\, \bra \!V_{\mu\nu} f^{\mu\nu}_+ \!\ket
+ \frac{i\, G_V}{2\sqrt{2}}\, \bra \! V_{\mu\nu} [u^\mu, u^\nu] \!\ket
\nonumber \\ &&\mbox{}
 + \frac{F_A}{2\sqrt{2}}\, \bra \!A_{\mu\nu} f^{\mu\nu}_- \!\ket
 + \sqrt{2}\, \lambda_1^{hA}\,  \partial_\mu h \, \bra \! A^{\mu \nu} u_\nu \!\ket\, .
\end{eqnarray}
%

\begin{figure}
\vspace*{-0.5cm}
\begin{center}
\psfrag{w}{a}
\includegraphics[scale=0.4]{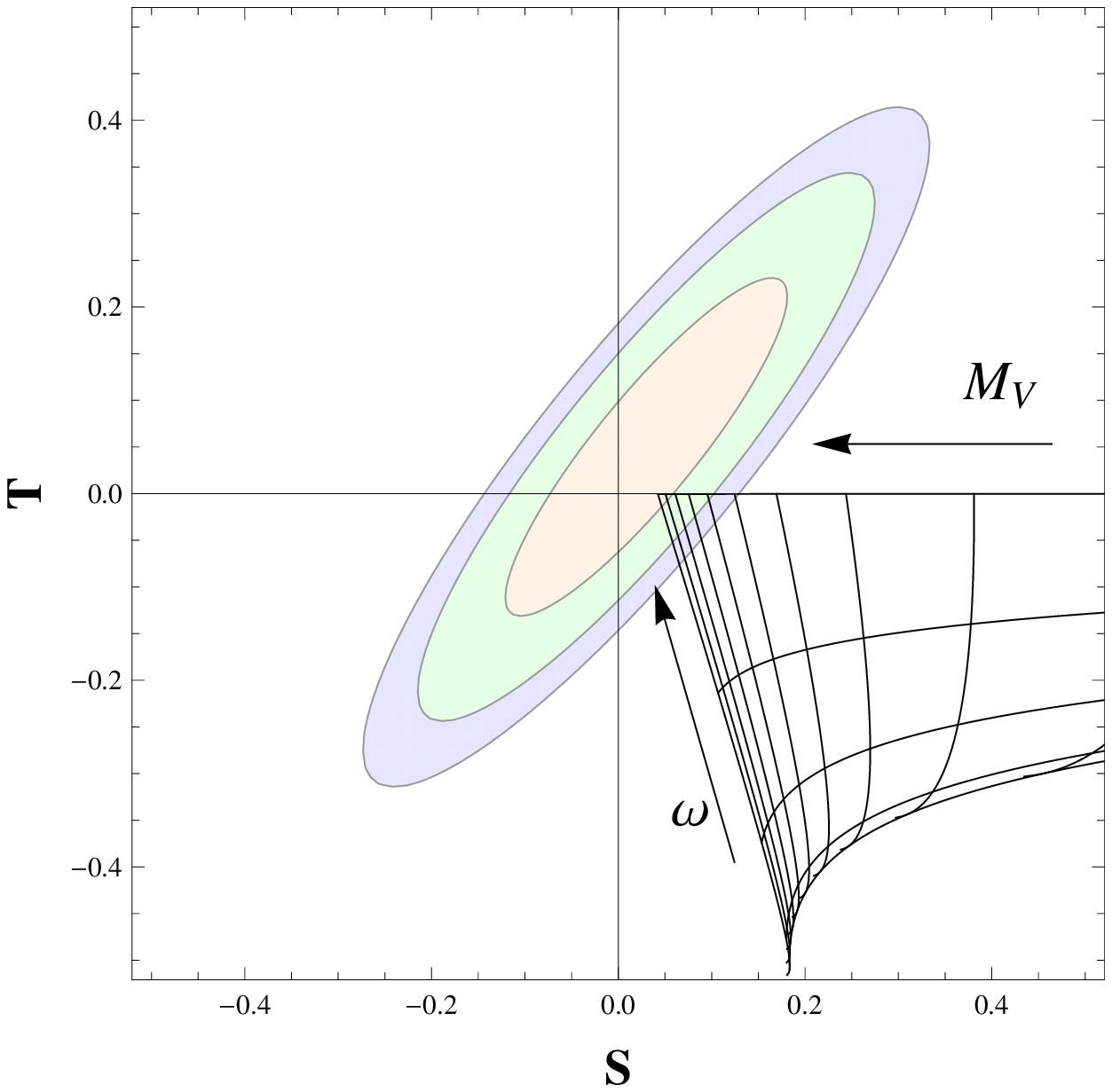}
\qquad
\includegraphics[scale=0.6]{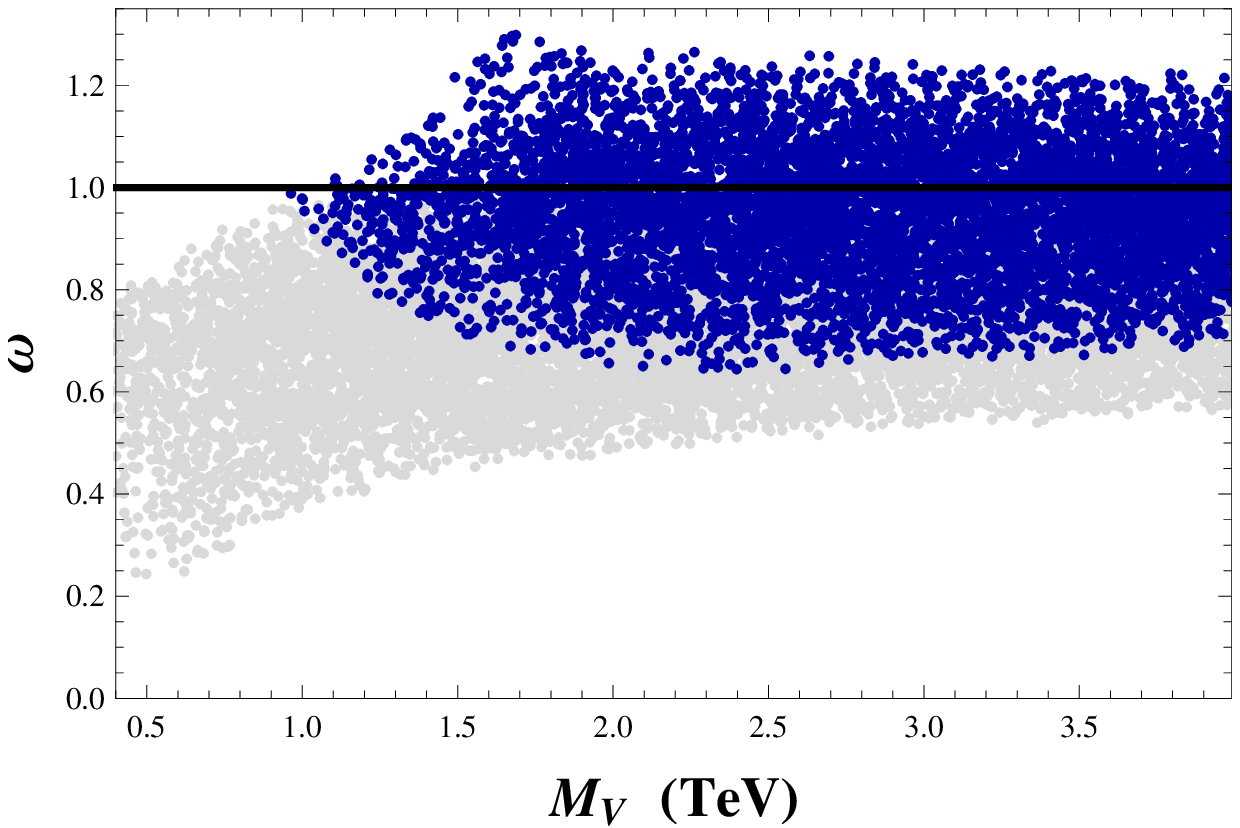}
\\
\vspace*{-0.75cm}
\flushleft\hspace*{5cm} {\bf a)} \hspace*{8cm} {\bf b)}
\vspace*{-0.25cm}
\caption{\small{
{\bf a)} NLO determinations of $S$ and $T$, imposing the two WSRs.
The approximately vertical curves correspond to constant values
of $M_V$, from $1.5$ to $6.0$~TeV at intervals of $0.5$~TeV.
The approximately horizontal curves have constant values
of $a$:
$0.00, \, 0.25, 0.50, 0.75, 1.00$.
The arrows indicate
the directions of growing  $M_V$ and $a$.
The ellipses give the experimentally allowed regions at 68\% (orange), 95\% (green)
and 99\% (blue) confidence level (CL).
{\bf b)}
Scatter plot for the 68\% CL region, in the case
when only the first WSR is assumed.
The dark blue and light gray regions
correspond, respectively,  to
$0.2<M_V/M_A<1$ and $0.02<M_V/M_A<0.2$.
}}
\label{fig.S+T}
\end{center}
\vspace*{-1cm}
\end{figure}

%
%
In order to compute  $S$ and $T$  up to the one-loop level
we use the dispersive representations~\cite{Peskin:92,S+T+scalar},
%
\begin{equation}
S\, =\, \Frac{16 \pi}{g^2\tan\theta_W}\,
\Int_0^\infty \, \Frac{{\rm dt}}{t} \, [\, \rho_S(t)\, - \, \rho_S(t)^{\rm SM} \, ]\, , \qquad
T \,=\,  \Frac{4 \pi}{g'^2 \cos^2\theta_W}\, \Int_0^\infty \,\Frac{{\rm dt}}{t^2} \,
[\, \rho_T(t) \, -\, \rho_T(t)^{\rm SM} \,] \, ,
\label{eq.sum-rules}
\end{equation}
with $\rho_S(t)\,\,$ the spectral function of the $W^3B$ correlator~\cite{S-Higgsless,Peskin:92}
and $\rho_T(t)\,\,$ the spectral function of the difference
of the neutral and charged Goldstone self-energies~\cite{S+T+scalar}.
The calculation of $T$ above has been simplified by means of the
Ward-Takahashi relation $T=Z^{(w^+)}/Z^{(w^0)} -1$ ~\cite{Barbieri:1992dq}.
%
Only the lightest two-particle cuts have been considered in $\rho_S(t)$ and $\rho_T(t)$,
respectively, $\{ww, w h\}$ and $\{Bw, Bh\}$.
%
%
Since $\rho_S(t)^{\rm SM}\stackrel{t\to\infty}{\longrightarrow} 0$,  the convergence of
the Peskin-Takeuchi sum-rule
requires  $\rho_S(t)\stackrel{t\to\infty}{\longrightarrow} 0$.
%
%
%
%
Furthermore, assuming that weak isospin and parity are good symmetries of the BSM strong dynamics,
the $W^3 B$ correlator is proportional to the difference of
the vector and axial-vector two-point Green's functions~\cite{Peskin:92}.
In asymptotically-free gauge theories this difference
vanishes at $s\to\infty$ as $1/s^3$ \cite{Bernard:1975cd},
implying the (tree-level) LO WSRs~\cite{WSR},
%
\begin{equation}
F_{V}^2 - F_{A}^2  = v^2\quad \mbox{(1st WSR), }   \qquad
F_{V}^2  \,M_{V}^2 - F_{A}^2 \, M_{A}^2  = 0\quad \mbox{(2nd WSR). }
\end{equation}
However, although the 1st WSR is expected to be true in gauge theories
with non-trivial ultraviolet fixed points~\cite{S-Orgogozo:11,WalkingTC},
the 2nd WSR is questionable in some of these models.
Thus, two alternative scenarios are studied in Ref.~\cite{S+T+scalar}:
one assuming the two WSRs
and another assuming just the 1st~WSR.
At tree-level one has the the LO determinations~\cite{Peskin:92,S+T+scalar,S-Higgsless}
\vspace*{-0.125cm}
\begin{eqnarray}
&& S_{\mathrm{LO}}
\; =\; 4\pi\, \left(\frac{F_V^2}{M_V^2}\, -\, \frac{F_A^2}{M_A^2} \right)
\; =\; \frac{4\pi v^2}{M_V^2}\,  \left( 1 + \frac{M_V^2}{M_A^2} \right)
\qquad\qquad  \mbox{(1st \& 2nd WSR)}
\, ,
\label{eq.S-LO}
\\
&&
S_{\mathrm{LO}}
= 4\pi \left\{ \frac{v^2}{M_V^2} + F_A^2 \left( \frac{1}{M_V^2} - \frac{1}{M_A^2} \right)
\right\}> \frac{4\pi v^2}{M_V^2}
\qquad\qquad  \mbox{(1st WSR \& }M_V<M_A \mbox{)}  .
\label{eq.LO-S+1WSR}
\nn
\end{eqnarray}
In the first case, the two WSRs imply $M_V<M_A$ and determine  $F_V$ and $F_A$
in terms of the resonance masses~\cite{Peskin:92,S+T+scalar,S-Higgsless,RChT}.
In the second case, it is not possible to extract a definite prediction with just the 1st WSR
but one can still derive the inequality above if one assumes a similar mass hierarchy $M_V<M_A$.
On the other hand, this inequality flips direction
if  $M_A<M_V$ or turns into an equality in the degenerate case $M_V=M_A$~\cite{S+T+scalar}.
At NLO the computed $W^3 B$ correlator
is given by the $ww$ and $hw$ cuts, whose contributions
to the $\rho_S(t)$ spectral function
would have an unphysical grow at high energies unless
$F_V G_V = v^2$ and $F_A \lambda^{hA}_1 = a v$~\cite{S+T+scalar,S-Higgsless,RChT}.
%
%
Thus, we obtain the NLO prediction~\cite{S+T+scalar}
\vspace*{-0.125cm}
\begin{eqnarray}
S  &=&   4 \pi v^2 \left(\frac{1}{M_{V}^2}+\frac{1}{M_{A}^2}\right)
\label{eq.S-NLO}
\label{eq.1+2WSR}
\\ &&
 + \frac{1}{12\pi}
\bigg[ \log\frac{M_V^2}{m_{H}^2}  -\frac{11}{6}
+\;\frac{M_V^2}{M_A^2}\log\frac{M_A^2}{M_V^2}
 - \frac{M_V^4}{M_A^4}\, \bigg(\log\frac{M_A^2}{m_{S_1}^2}-\frac{11}{6}\bigg) \bigg]
\qquad \mbox{(1st \& 2nd WSR)\,,}
\nn\\
S &>&   \frac{4 \pi v^2}{M_{V}^2} + \frac{1}{12\pi}  \bigg[ \log\frac{M_V^2}{m_{H}^2}
-\frac{11}{6}
- a^2 \bigg(\!\log\frac{M_A^2}{m_{S_1}^2}-\frac{17}{6}
 + \frac{M_A^2}{M_V^2}\!\bigg) \bigg]  ,
\label{eq.lower-bound-1WSR}
\qquad   \mbox{(1st WSR \& }M_V<M_A \mbox{).}
\nn
\end{eqnarray}
In the two-WSRs scenario, in order to enforce the 2nd WSR at NLO
one needs the additional constraint
$a = M_V^2/M_A^2$ (hence restricted to the range $0\leq a \leq 1$).
Again, the inequality in the last line flips direction or
turns into an equality when, respectively, $M_A<M_V$ or $M_V=M_A$.

\begin{table}[!t]
\vspace*{-0.5cm}\begin{center}
\begin{tabular}{ c|c|c }
 & $a$ &  $M_V$
\\[5pt] \hline
\rule{0pt}{3ex}
two WSRs & $0.97\,$--$\, 1$ &  $>5$~TeV
\\[5pt] \hline
\rule{0pt}{3ex}
Only 1st WSR: $\quad  0.2<M_V/M_A<1$     & $0.6\,$--$\, 1.3$ &  $>1$~TeV
\\[5pt] 
\rule{0pt}{3ex}
$\qquad\qquad\qquad\quad   0.5<M_V/M_A<1$     & $0.84\,$--$\, 1.30$ &  $>1.5$~TeV
\\[5pt] 
\rule{0pt}{3ex}
$\qquad\qquad\qquad  M_V/M_A=1$     & $0.97\,$--$\, 1.30$ &  $>1.8$~TeV
\\[5pt] 
\rule{0pt}{3ex}
$\qquad (M_V>1$~TeV$)^\dagger  \quad  1<M_V/M_A<2$     & $0.7\,$--$\, 1.9$ &   $>1$~TeV$^\dagger$
\\[5pt] \hline
\end{tabular}
\caption{\small
Allowed range for the $M_V$ and $a$  at the 68\% CL for the two-WSRs
(where $V$ and $A$ are very degenerate since $M_V^2/M_A^2=a$ in this case)
and only-1st-WSR cases
(for various values $M_V/M_A$).
In the last line we also impose the restriction$^\dagger$ $M_V>1$~TeV.
}
\label{tab.S+T}
\end{center}
\vspace*{-1.25cm}
\end{table}

At LO, $\rho_T(t)$
is zero and one has  $T_{\rm LO}=0$.
At NLO, where we enforce the $\rho_S(t)$
constraints $F_V G_V = v^2$ and $F_A \lambda^{hA}_1 = a v$,
we find that $\rho_T(t)\stackrel{t\to\infty}{\longrightarrow}0$
and obtain the NLO prediction
\vspace*{-0.25cm}
\begin{equation}
 T\; =\;  \frac{3}{16\pi \cos^2 \theta_W} \bigg[ 1 + \log \frac{m_{h}^2}{M_V^2}
 - a^2 \left( 1 + \log \frac{m_{h}^2}{M_A^2} \right)  \bigg]  \, ,
\label{eq:T}
\end{equation}
%



In Fig.~\ref{fig.S+T},
we show the compatibility between the experimental determinations
for $S$ and $T$~\cite{S+T-phenomenology}
and our NLO determinations in both scenarios.
The numerical results in Table~\ref{tab.S+T} show that
the precision electroweak data
requires resonance masses over the TeV and
the $hWW$ coupling to be close to the
SM one ($a^{\rm SM}=1$),
in agreement with present LHC bounds~\cite{LHC-exp}.

To conclude, we emphasize that, remarkably, just by considering
the experimental $m_h$ (the only LHC input)
and the EW precision observables (LEP input), the allowed region
concentrates around $a\simeq 1$ for reasonable values of the splitting
$M_V/M_A\sim \cO(1)$   (see Fig.~\ref{fig.S+T} and Table~\ref{tab.S+T}).

%
%
%

\end{document}